\documentclass[reprint,aps,prl,showpacs]{revtex4-1}
\usepackage{physics}
\usepackage{amsmath}
\usepackage{graphicx}
\usepackage{verbatim}
\usepackage{color}
\usepackage{subfigure}
\usepackage{hyperref}
\usepackage{natbib}
\usepackage{placeins}
\usepackage{adjustbox}
\usepackage{soul}

\begin{document}
\title{Microcanonical entropy: consistency and adiabatic invariance}
\author{Arash Tavassoli}
\author{Afshin Montakhab}
\email{montakhab@shirazu.ac.ir}
\affiliation{Department of Physics, College of Sciences, Shiraz University, Shiraz 71946-84795, Iran}

\begin{abstract}
Attempts to establish microcanonical entropy as an adiabatic invariant date back to works of Gibbs and Hertz. More recently, a consistency relation based on adiabatic invariance has been used to argue for the validity of Gibbs (volume) entropy over Boltzmann (surface) entropy.  Such consistency relation equates derivatives of thermodynamic entropy to ensemble average of the corresponding quantity in micro-state space (phase space or Hilbert space). In this work we propose to re-examine such a consistency relation when the number of particles ($N$) is considered as the independent thermodynamic variable. In other words, we investigate the consistency relation for the chemical potential which is a fundamental thermodynamic quantity. We show both by simple analytical calculations as well as model example that \emph{neither} definitions of entropy satisfy the consistency condition when one considers such a relation for the chemical potential. This remains true regardless of the system size. Therefore, our results cast doubt on the validity of the adiabatic invariance as a required property of thermodynamic entropy. We close by providing commentary on the derivation of thermostatistics from mechanics which typically leads to controversial  and inconsistent results.
\end{abstract}
\pacs{05.70.-a, 05.30.-d, 05.90.+m, 01.70.+w}
\maketitle

\emph{Introduction:} In the past one and half century, since the advent of principles of statistical mechanics, the tools of statistical thermophysics have been extremely successful in explaining the macroscopic properties of a wide range of systems\cite{gibbs1902elementary,4_boltzmann_1964,pathria1996statistical,huang1987statistical,swendsen2012an}. The success of such theories is crucially dependent on the principle of maximum entropy\cite{callen2006thermodynamics}, which clearly requires a definition of the fundamental quantity called entropy (S). The typical approach is to consider a closed (microcanonical) system which obeys conservation of energy and therefore facilitates application of the laws of mechanics (quantum or classical). Thermostatistical laws for more realistic open systems are then derived using entropy definition, developed for microcanonical system. For this basic reason, microcanonical entropy is considered as the most fundamental quantity in statistical mechanics. Despite the widespread success of the general theory, the fundamental definition of entropy has proven difficult to decide, as various definitions of entropy are used by various authors\cite{gibbs1902elementary,4_boltzmann_1964,tsallis2009introduction,shannon2015mathematical}.

Likewise, ``foundations of statistical mechanics" has encountered many theoretical challenges which have caused much confusion and, despite many important contributions by great scientists over the past century, have failed to be fully resolved or clarified. Chief among such problems are the apparent paradox between microscopic reversibility and macroscopic irreversibility \cite{boltzmann1896entgegnung,swendsen2012an}, ergodic problem \cite{khinchin1949mathematical}, and the problem of exorcising Maxwell's demon \cite{harvey2002maxwell}. The common aspect of all such problems is the belief (generally accepted by most physicists) that macroscopic thermostatistical laws must be directly derived from the laws of mechanics governing many particle systems.  This essentially \emph{reductionist} point of view has been challenged by recent studies of complex systems where one has come to realize that ``more is different" \cite{anderson1972more,mazenko2000equilibrium}.

More recently, there has been another controversy which has attracted much attention \cite{dunkel2014consistent,frenkel2015gibbs,vilar2014communication,schneider2014comment,dunkel2014reply,dunkel2014reply2,campisi2015construction,hilbert2014thermodynamic,swendsen2014negative,swendsen2015gibbs,swendsen2015continuity,hanggi2015meaning,buonsante2015phase,
anghel2015stumbling,cerino2015consistent,wang2015critique} over what is referred to as ``consistent thermostatistics"\cite{dunkel2014consistent}. In fact such a consistent formulation of thermostatistics is again based on mechanical views which formulates microcanonical entropy as a mechanical adiabatic invariant. This view has its roots in the original works of Gibbs\cite{gibbs1902elementary} and Hertz\cite{hertz1910mechanischen}. Ever since these early works, many authors have advocated that microcanonical entropy be an adiabatic invariant since it offers many mathematical advantages and strengthens the mechanical foundations of thermostatistics\cite{campisi2005mechanical}. More recently\cite{dunkel2014consistent}, it has been shown that the condition of adiabatic invariance of entropy (as well as ergodicity), imposes a consistency condition which favors Gibbs (volume) entropy over the frequently used and generally accepted Boltzmann (surface) entropy.  This is so, simply because, it is believed that Gibbs entropy ($S_G$) is an adiabatic invariant while Boltzmann entropy ($S_B$) is not(unless thermodynamic limit is imposed). One immediate consequence of such a condition is that ``consistent thermostatistics forbids negative absolute temperatures" \cite{dunkel2014consistent}. In this Letter we propose to inspect the consistency condition of microcanonical entropy more closely. In particular, we investigate the consistency condition for chemical potential by considering the number of particles ($N$) as the independent thermodynamic variable. Our results prove that under such scenario the consistency condition fails to be valid for either entropies, thus proving that neither entropies are adiabatic invariants. We note that our results essentially pose a challenge to $S_G$ as it is the microcanonical entropy which is thought to be an adiabatic invariant and thus consistent. $S_B$, on the other hand, is not generally believed to be an adiabatic invariant (and thus not consistent) for at least finite systems. However, more importantly, our results pose yet another challenge to the traditional views of entropy based on laws of mechanics.

\emph{Microcanonical entropy and consistency relation:} Microcanonical entropy definitions are based on identification of micro-state space and their subsequent multiplicities. Gibbs and Boltzmann entropies are under special considerations. These entropy definitions are:
\begin{align}
&S_B=k \ln \epsilon\omega ;\; &\omega=\Tr [\delta(E-H)]\\
&S_G=k \ln \Omega ;\; &\Omega=\Tr [\Theta(E-H)]\label{SGdef}
\end{align}
Here $\delta$ is the Dirac delta function, $\Theta$ is the Heaviside step function, $k$ is the Boltzmann constant, $\epsilon$ is a constant with energy dimension\footnote{The Boltzmann entropy is often defined as the famous equation $S_B=k \ln W$, where $W$ is the degeneracy of system's internal energy. The adoption of $\epsilon \omega$ instead of $W$ is due to some conceptual issues (see SI of Ref.~\onlinecite{dunkel2014consistent}). This distinction has no bearing on the results presented here.} and $E$ and $H$  are the energy and Hamiltonian of microcanonical system, respectively. Boltzmann entropy is preferred by essentially  all authors of standard statistical mechanic textbooks\cite{pathria1996statistical,huang1987statistical,swendsen2012an,callen2006thermodynamics}. Gibbs entropy, however, has been used by various authors due to its historical as well as (certain) mathematical advantages\cite{khinchin1949mathematical,berdichevsky1997thermodynamics}. Although in many macroscopic systems the distinction between such entropy definitions becomes irrelevant, in certain systems such as ones with bounded energy spectrum, they lead to distinctly different thermodynamic properties\cite{dunkel2014consistent,vilar2014communication,frenkel2015gibbs}. For example, if one considers the microcanonical definition of temperature, $T\equiv (\pdv{S}{E})^{-1}$, the Boltzmann and Gibbs temperatures are given by   $T_B=\frac{\omega}{k\omega'}$ and $T_G=\frac{\Omega}{k\omega}$, respectively. Then, in the population inverted regime of such systems $\omega'\equiv \pdv{\omega}{E} <0$ and consequently $T_B<0$\cite{ramsey1956thermodynamics,klein1956negative,rapp2010equilibration,braun2013negative}. In contrast, $T_G$ is always a positive quantity\cite{dunkel2014consistent,berdichevsky1991negative}. Recently a consistency condition has been employed in order to distinguish between the two definitions of entropy\cite{dunkel2014consistent}. The consistency condition, on one hand assumes a locally invertible entropy function,  $S(E,A_\mu)\leftrightarrow E(S,A_\mu)$, where $A_\mu$'s are various thermodynamic variables such as volume (V), particle number (N) or magnetic field (B), leading to $T\pdv{S}{A_\mu}=- \pdv{E}{A_\mu}$,  while on the other hand relies on micro-state (ensemble) averaging to calculate that $\pdv{E}{A_\mu}=\expval{\pdv{H}{A_\mu}}$, which then leads to:
\begin{equation}
T\pdv{S}{A_\mu}=-\expval{\pdv{H}{A_\mu}}
\label{consistency}
\end{equation}

Such a relation is called consistency relation since it equates various thermodynamic quantities on the left hand side to their micro-state (ensemble) averages on the right hand side. One can show that consistency relation is satisfied when $S$ is an adiabatic invariant.\cite{dunkel2014consistent,campisi2005mechanical}.

It is claimed \cite{dunkel2014consistent,campisi2005mechanical} that $S_G$ respects the consistency relation (Eq.~\ref{consistency}) while $S_B$ does not satisfy it under general conditions.  However, a proof has been provided to justify the consistency relation for Boltzmann entropy in the thermodynamic limit\cite{frenkel2015gibbs}. This is already clear since $S_B\rightarrow S_G$ as $N\rightarrow \infty$ for typical systems. However, for systems with bounded energy, this convergence is violated and the consistency of Boltzmann entropy deserves to be re-examined\cite{campisi2015construction}. In addition, there is an ongoing debate as to the necessity of the thermodynamic limit, as regards to the validity of thermodynamic laws\cite{berdichevsky1997thermodynamics,dunkel2014consistent,hilbert2014thermodynamic,hanggi2015meaning,frenkel2015gibbs,swendsen2014negative,swendsen2015gibbs}.

It is important to note that the consistency condition of Eq.~(\ref{consistency}), is an essential ingredient in derivation of thermodynamics from the laws of mechanics\cite{berdichevsky1997thermodynamics,khinchin1949mathematical}. This is so because Eq.~(\ref{consistency}), is the requirement that thermodynamic entropy of a system be a mechanical adiabatic invariant. In the discussions that have ensued since the publication of Ref.~[\onlinecite{dunkel2014consistent}], the consistency relation, Eq.~(\ref{consistency}), has been used where $A_\mu$ is a general thermodynamic variable like magnetic field strength\cite{dunkel2014consistent,campisi2015construction,hilbert2014thermodynamic,frenkel2015gibbs}. However, chemical potential is a fundamental thermodynamic quantity whose consistency should be checked by replacing $A_\mu=N$ in Eq.~(\ref{consistency}). In other words, which definition of thermodynamic chemical potential is consistent with its microcanonical statistical mechanical counterpart? We propose to investigate this question in what follows. Surprisingly, our results indicate that neither definition of thermodynamic entropy (Gibbs nor Boltzmann) satisfies Eq.~(\ref{consistency}) when $A_\mu=N$. This simply follows because in order to prove Eq.~(\ref{consistency}) one is required to move the derivative inside the integral (or trace, see Eq.~\ref{SGdef}) which is clearly not allowed when $A_\mu=N$ since the integration on phase space volume is implicitly $N$ dependent. To further demonstrate our point, we will prove that the assumption of the validity of Eq.~(\ref{consistency}) for $S_G$ will lead to wrong results by a simple analytic argument. We also provide an explicit calculation where we will define the absolute value of the difference of the two sides of Eq.~(\ref{consistency}) as a measure of ``inconsistency" and show that for a system of $N$ (finite or otherwise) harmonic oscillators, it remains a finite \emph{nonzero} value for both definitions of entropy. We therefore prove that \emph{neither} definitions of entropy are consistent whether they are considered in a finite system \emph{or} in the thermodynamic limit.

\emph{Inconsistency of microcanonical entropy definitions:} An important thermodynamic parameter of any system is the particle number $N$. Consistency in conjunction with the necessity of any thermodynamic entropy to be an adiabatic invariant in the reversible processes in which the system's particles number is subjected to quasi-static change, implies the necessity of the following equation:
\begin{equation}
T\pdv{S}{N}\overset{?}{=}-\expval{\pdv{H}{N}}
\label{Nconsistency}
\end{equation}
In spite of the fact that $N$ is a discrete variable we can consider it as a continuous one if $N>>1$  as is done in standard practice\cite{pathria1996statistical,huang1987statistical,callen2006thermodynamics}.

It is worthwhile to review the mathematical proof of the Gibbs entropy's consistency for a general thermodynamic parameter $A_\mu$. One may write\cite{dunkel2014consistent}:
\begin{align}
T_G\pdv{S_G}{A_\mu}=&\frac{\pderivative{A_\mu}\Tr[\Theta (E-H)]}{\omega} \label{proof1}\\
=& \frac{\Tr[\pderivative{A_\mu}\Theta (E-H)]}{\omega}\label{proof2}\\
=&-\Tr[\frac{\pdv{H}{A_\mu}\delta (E-H)}{\omega}]=-\expval{\pdv{H}{A_\mu}}\label{proof3}
\end{align}

In a classical mechanical notation one may swap ``$\Tr$" with ``$\underbrace{\idotsint}_{6N}\frac{\dd q^{3N}\dd p^{3N}}{h^{3N}}$"\cite{campisi2005mechanical}. However, the validity of the above relation is in doubt for the important parameter $A_\mu=N$ because the interchange of integral with derivative (which is a necessary step to conclude Eq.~(\ref{proof2}) from Eq.~(\ref{proof1})) is not justified when the (number of) integrated variables appears as a derivative. In other words, for the particular case $N$, one cannot simply move the derivative inside the integral. In the context of classical mechanics, the reason is the dependence of the number of phase space's dimensions on $N$ in addition to the dependence of Hamiltonian ($H$) on this parameter.

This can clearly be shown for systems with bounded energy by assuming the opposite. Assume that Eq.~(\ref{Nconsistency}) holds for $\lbrace T,S \rbrace=\lbrace T_G,S_G \rbrace$ in a system with arbitrary energy $E$ in the range $(0,E_{max})$. Eq.~(\ref{Nconsistency}) then implies:
\begin{equation}
\pderivative{N}\Tr [\Theta (E-H)]=\Tr [\pderivative{N} \Theta(E-H)]
\label{disprove1}
\end{equation}
from which it is easy to deduce:
\begin{equation}
\pderivative{N}\Tr [\Theta (H-E)]=\Tr [\pderivative{N} \Theta(H-E)]
\label{disprove2}
\end{equation}
By adding the left hand sides of Eq.~(\ref{disprove1}) and Eq.~(\ref{disprove2}) we get:
\begin{align}
\pderivative{N}\Tr [\Theta (E-H)]+&\pderivative{N}\Tr [\Theta (H-E)]\nonumber\\
=&\pderivative{N}\Tr [I] \neq 0
\label{disprove3}
\end{align}
Here $I$ is the identity matrix and $\Tr[I]$ means the total number of all possible states which of course is a function of $N$.  The contradiction arises when we add the right hand sides of Eq.~(\ref{disprove1}) and Eq.~(\ref{disprove2}) because:
\begin{align}
\Tr &\qty[\pderivative{N}\Theta (E-H)]+\Tr [\pderivative{N}\Theta (H-E)]\nonumber\\
=&\Tr {[-\delta (E-H)+\delta (H-E)]\pdv{H}{N}}= 0
\label{disprove4}
\end{align}

We have clearly arrived at a contradiction assuming the validity of Eq.~(\ref{Nconsistency}). Therefore for systems with bounded energy:
\begin{equation}
T_G\pdv{S_G}{N}\neq -\expval{\pdv{H}{N}}
\end{equation}
This means that Gibbs entropy is not a global adiabatic invariant (consistent) definition of entropy.

\emph{A system of simple harmonic oscillators:}\label{SHOinconsistency}
As a concrete example of inconsistency, we propose to calculate both sides of Eq.~(\ref{Nconsistency}) for Boltzmann and Gibbs entropy for a system consisting of $N$ independent one dimensional quantum harmonic oscillators.

Consider a quantum system consisting of $N$ simple harmonic oscillators with microcanonical energy $E$. Employing  $\hbar\omega$ as the unit of energy, the eigenvalues of Hamiltonian of this system are:
\begin{equation}
\mathcal{E} =n_1+n_2+...+n_N+\frac{N}{2}
\label{Heigenval}
\end{equation}
Where $n_i$ is the quantum number of i'th oscillator. Furthermore, the degeneracy of energy $\mathcal{E} $ is easily computed as:
\begin{equation}
W(\mathcal{E} )=\frac{(\mathcal{E} +\frac{N}{2}-1)!}{(N-1)!(\mathcal{E} -\frac{N}{2})!}
\label{HW}
\end{equation}
First we compute the right hand side of Eq.~(\ref{Nconsistency}):
\begin{align}
\expval{\frac{\delta H}{\delta N}}=&\frac{1}{W} \sum_{n_1,n_2,...,n_N} \bra{n_1,n_2,...,n_N}\nonumber\\
\times&\frac{\delta H}{\delta N}\delta_{HE} \ket{n_1,n_2,...,n_N}\nonumber\\
=&\frac{1}{W}\sum_{n_1,n_2,...,n_N}\bra{n_1,n_2,...n_N}\nonumber\\
\times& (H_N-H_{N-1}) \delta_{HE}\ket{n_1,n_2,...n_N}\nonumber\\
=&\frac{1}{W}\sum_{n_1+...+n_N+\frac{N}{2}=E}(\mathcal{E} _N-\mathcal{E} _{N-1})\nonumber\\
=&\frac{1}{W}\sum_{n_1+...+n_N+\frac{N}{2}=E}(n_N+\frac{1}{2})
\label{H1}
\end{align}
Note that $\delta_{HE}$ is the Kronecker's delta which appears here regardless of entropy definition, since here we are performing ensemble averaging over \emph{accessible} states.

In order to simplify the above, we define $g(x)$ as the total number of states which satisfy the constraint $n_1+n_2+...+n_{N-1}=x$. We note that $g(x)=\frac{(x+N-2)!}{(N-2)!x!}$. Therefore:
\begin{align}
\expval{\frac{\delta H}{\delta N}}=&\frac{1}{W} \sum _{n_1+n_2 +...+n_N+\frac{N}{2}=E}(n_N+\frac{1}{2})\nonumber\\
=&\frac{1}{W}\lbrace(0+\frac{1}{2})g(E-\frac{N}{2}-0)\nonumber\\
+&(1+\frac{1}{2})g(E-\frac{N}{2}-1)\nonumber\\+&...
+(E-\frac{N}{2}+\frac{1}{2})g(E-\frac{N}{2}-(E-\frac{N}{2})\rbrace\nonumber\\
=&(\frac{1}{W})\sum_{i=0}^{E-\frac{N}{2}} (i+\frac{1}{2})\frac{(E+\frac{N}{2}-i-2)!}{(E-\frac{N}{2}-i)!(N-2)!}\nonumber\\
=&\frac{E}{N}
\label{H2}
\end{align}
 In order to arrive at the last step we have employed the use of symbolic summation. Note that this result is independent of the definition of entropy.

Now, for Boltzmann entropy the left hand side of Eq.~(\ref{Nconsistency}) is easily computed using Eq.~(\ref{HW}) in conjunction with the Sterling approximation:
\begin{align}
T_B\pdv{S_B}{N}\approx
\frac{\ln (\text{N}-1)-\ln \left(\text{E}-\frac{\text{N}}{2}\right)}{\ln \left(\text{E}-\frac{\text{N}}{2}\right)-\ln
   \left(\text{E}+\frac{\text{N}}{2}-1\right)}+\frac{1}{2}
\end{align}
We therefore define ``inconsistency" as the absolute value of the difference between two sides of Eq.~(\ref{Nconsistency}):
\begin{figure}
   \subfigure{\includegraphics[width=6cm,height=4cm]{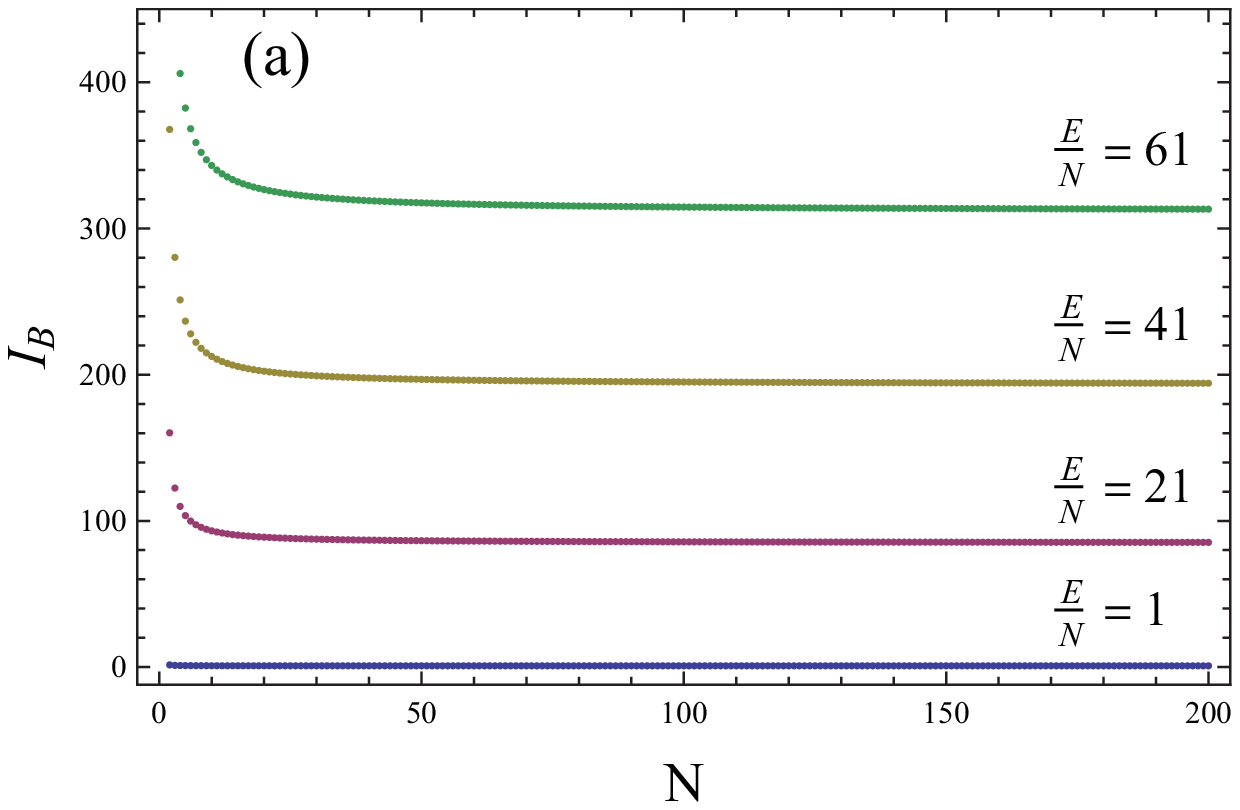}\label{IB}}\quad
    \subfigure{\includegraphics[width=6cm,height=4cm]{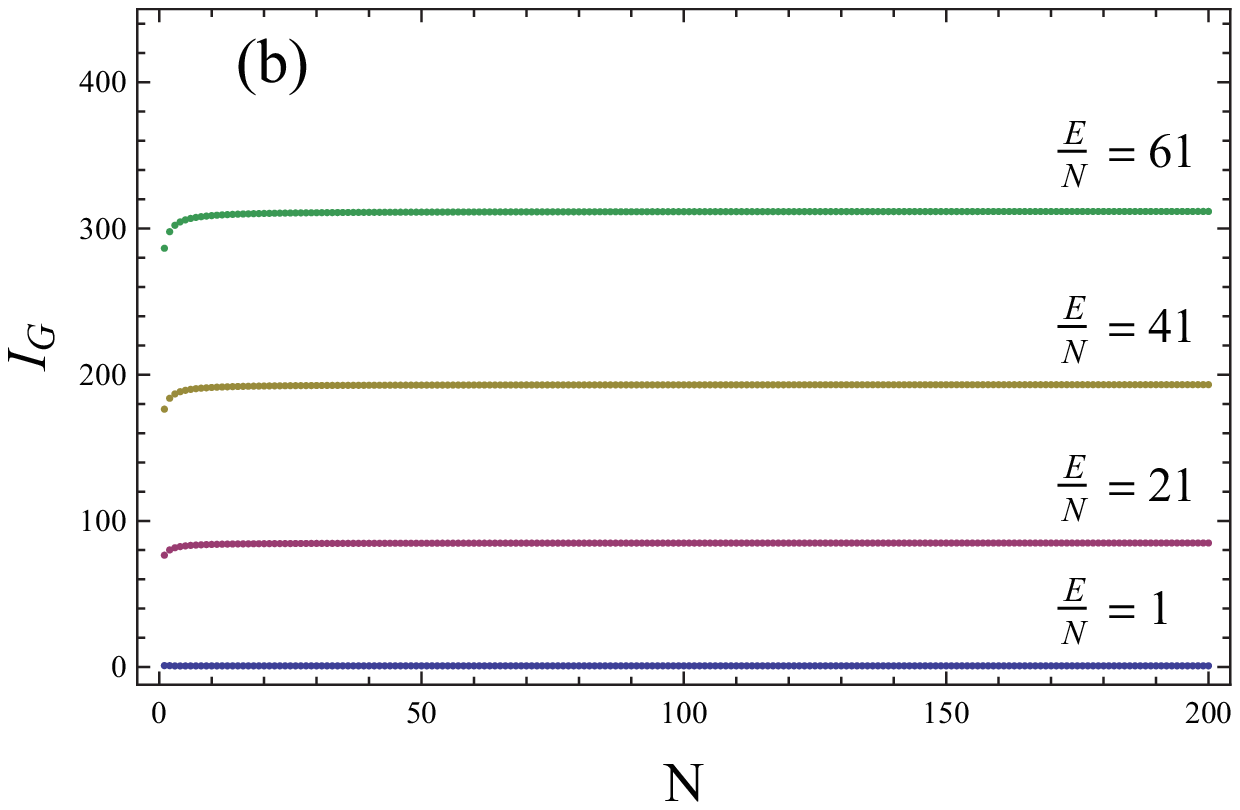}\label{IG}}
\caption{(Color online) Inconsistency vs.~$N$ for a system of simple harmonic oscillators, for different energy densities ($\frac{E}{N}$). (a) Boltzmann inconsistency($I_B$) (b) Gibbs inconsistency($I_G$). }
\end{figure}
\begin{equation}
I\equiv \abs{T\pdv{S}{N}+\expval{\pdv{H}{N}}}
\end{equation}
which leads to:
\begin{align}
&I_B= \abs{T_B\pdv{S_B}{N}+\expval{\pdv{H}{N}}}\approx \nonumber\\
&\abs{\frac{\ln (\text{N}-1)-\ln \left(\text{E}-\frac{\text{N}}{2}\right)}{\ln \left(\text{E}-\frac{\text{N}}{2}\right)-\ln
   \left(\text{E}+\frac{\text{N}}{2}-1\right)}+\frac{1}{2}+\frac{E}{N}}
\end{align}
Clearly $S_B$ is not consistent with respect to $N$ in the thermodynamic limit. Fig.~\ref{IB} shows results of computing exact $I_B$ (without the Sterling approximation) up to $N=200$.

In order to calculate inconsistency for the Gibbs entropy, $I_G$, one can numerically integrate Eq.~(\ref{HW}) in order to calculate $S_G$. The results for $I_G$ are shown in Fig.~\ref{IG}. It can be seen that $I_B$ and $I_G$ quickly saturate to a constant value as $N$ increases, showing considerable inconsistency for both $S_B$ and $S_G$ regardless of the actual value of $N$. It is interesting to note that $I_B\approx I_G$ for large $N$.

We have therefore shown that a model of simple harmonic oscillators exhibits considerable inconsistency regardless of entropy definition (Gibbs or Boltzmann) or system size (finite or infinite).

\emph{Conclusions:}
Recently, much attention has been given to a ``consistent" thermostatistics formalism\cite{dunkel2014consistent,campisi2015construction,hilbert2014thermodynamic,hanggi2015meaning,frenkel2015gibbs,swendsen2014negative,schneider2014comment}. Such an approach is motivated by making a direct connection between thermodynamic quantities on one hand to micro-state (ensemble) averages on the other hand.  For a closed system the consistency criterion,  Eq.~\ref{consistency}, is a direct consequence of the requirement that thermodynamic entropy be an adiabatic invariant. This criterion has been used to make detailed arguments in favor of one definition of entropy (Gibbs) vs.~another (Boltzmann)\cite{dunkel2014consistent,berdichevsky1991negative,campisi2005mechanical,campisi2015construction,hilbert2014thermodynamic,hanggi2015meaning,hertz1910mechanischen}. On the other hand, others have argued for consistency of Boltzmann entropy under specific conditions\cite{frenkel2015gibbs}. However, we have shown that neither entropy definitions satisfy the consistency requirement when one considers $N$ as the independent thermodynamic variable. We have shown this both by simple analytical calculations as well as for a model system of $N$ simple harmonic oscillators. Although we have shown inconsistency for both $S_B$ and $S_G$, it is $S_G$ that is believed to be a general adiabatic invariant and thus a candidate for consistent thermostatistics. One can draw various conclusions from the results we have presented here, but the most direct (and in our view the most important)  conclusion is that the long-held view of entropy as a mechanical adiabatic invariant is ill-founded.

 Consequently, it is worthwhile to consider the meaning of ``consistent thermostatistics". The consistency condition is a direct consequence of a (long-standing) attempt to derive the laws of thermodynamics from the laws of mechanics. These attempts have provided much controversy as well as various mathematical challenges (e.g.~ergodic theorem). Over the decades, such attempts have failed to provide a clear picture as to how one can derive thermodynamics from the laws of mechanics. The most famous inconsistency in this regard is the mechanical (time) reversibility vs.~thermodynamic irreversibility, which has been debated ever since Boltzmann's H-theorem and \emph{stosszahlans$\ddot{a}$tz}\cite{boltzmann1896entgegnung,huang1987statistical}.  Here, we have shown that \emph{any attempt to construct an entropy function as an adiabatic invariant leads to inconsistency in chemical potential.} We have therefore provided another example of inconsistent results when one attempts to construct thermodynamics based on the laws of mechanics.

The above-mentioned difficulties in deriving macroscopic laws from microscopic (mechanical) laws may have their roots in limitations of reductionism. While no one doubts that macroscopic properties of matter are results of ``particles and their interactions", attempts to model consciousness as a property of atoms and their interactions is patently a wrong start.  Similarly, while it is true that entropy is a property of closed interacting many-particle systems, traditional attempts to establish such direct connections have led to much controversy and confusion.  On the other hand, if one begins with the realization of limitations of mechanical approach (e.g. chaos) to describe large complex systems, one is naturally led to a purely probabilistic theory which attempts to establish relations among macroscopic parameters (e.g. pressure as a function of temperature) based on some very general and reasonable assumptions without resort to the laws of mechanics. Such a formulation is the information theoretic approach to thermostatistics \cite{jaynes1957information,jaynes1957information2,robertson1993statistical,mazenko2000equilibrium}. In such an approach one equates information entropy \cite{shannon2015mathematical}, which has no mechanical analog, to thermodynamic entropy. This approach, which is favored by the present authors, leads naturally to Boltzmann entropy for closed (microcanonical) system in accordance with standard formulations of statistical mechanics, but devoid of its conceptual difficulties.

Finally, we close by emphasizing that our results (lack of validity of Eq.~(4)) should not be interpreted as a way to deny the validity of well-established definitions of entropy, but as an objection to define entropy on a purely mechanical basis such as an adiabatic invariant.

\FloatBarrier

\bibliography{bib}

\end{document}